\documentclass[aps,prb,twocolumn, superscriptaddress, showpacs, a4paper, 10pt]{revtex4-1}
\usepackage [dvips]{graphicx}
\usepackage{graphicx,color}
\usepackage{ulem}
\usepackage{amsmath}
\usepackage{bm}
\usepackage{dcolumn}
\usepackage{multirow}
\usepackage{booktabs}
\usepackage[
	bookmarks=true,
	colorlinks=true,
	urlcolor=blue,
	linkcolor=blue,
	citecolor=blue,
	pdfpagemode=UseNone,
	pdfstartview=FitH,
	pdfencoding=auto
	]{hyperref}
\usepackage[all]{hypcap}
\begin{document}

\title{The role of electron confinement in Pd films for the oscillatory magnetic anisotropy in an adjacent Co layer}

\author{Sujit~Manna}
\email{smanna@physnet.uni-hamburg.de}
\affiliation{Max-Planck-Institut f\"ur Mikrostrukturphysik, Weinberg 2, D-06120 Halle, Germany}
\author{M. Przybylski}
\affiliation{Max-Planck-Institut f\"ur Mikrostrukturphysik, Weinberg 2, D-06120 Halle, Germany}
\affiliation{Academic Centre of Materials and Nanotechnology, and Faculty of Physics and Applied Computer Science, AGH University of Science and Technology, al. Mickiewicza 30, 30-059 Krak\'{o}w, Poland}
\author{D. Sander}
\affiliation{Max-Planck-Institut f\"ur Mikrostrukturphysik, Weinberg 2, D-06120 Halle, Germany}
\author {J.~Kirschner}
\affiliation{Max-Planck-Institut f\"ur Mikrostrukturphysik,
Weinberg 2, D-06120 Halle, Germany} \affiliation{Institut f\"ur
Physik, Martin-Luther-Universit\"at Halle-Wittenberg, D-06099
Halle, Germany}


\begin{abstract}
We demonstrate the interplay between quantum well states in Pd and the magnetic anisotropy in Pd/Co/Cu(001) by combined scanning tunneling spectroscopy (STS) and magneto optical Kerr effect (MOKE) measurements. Low temperature scanning tunneling spectroscopy reveals occupied and unoccupied quantum well states (QWS) in atomically flat Pd films on Co/Cu(001). These states give rise to sharp peaks in the differential conductance spectra. A quantitative analysis of the spectra reveals the electronic dispersion of the Pd(001) d-band ($\Delta$$_5$-type) along the $\Gamma$-X direction. \textit{In-situ} MOKE experiments on Pd/Co/Cu(1, 1, 13) uncover a periodic variation of the in-plane uniaxial magnetic anisotropy as a function of Pd thickness with a period of 6 atomic layers Pd. STS shows that QWS in Pd cross the Fermi level with the same periodicity of 6 atomic layers. Backed by previous theoretical work we ascribe the variation of the magnetic anisotropy in Co to QWS in the Pd overlayer. Our results suggest a novel venue towards  tailoring uniaxial magnetic anisotropy of ferromagnetic films by exploiting QWS in an adjacent material with large spin-orbit coupling.

\end{abstract}


\maketitle
\section{I. Introduction}
Tailoring magnetic anisotropy at the atomic scale in a magnetic nanostructures is one of the most important challenges in spintronics~\cite{Misiorny2013}. Systems with alternating ferromagnetic and noble metal layers have been studied intensely, as these structures may exhibit strong perpendicular anisotropy~\cite{Bland2005}. This is usually achieved by electronic hybridization between the valence levels of 3d transition metals (Fe, Co) with large magnetic moments, but relatively small spin-orbit coupling, and the valence level of 4d, 5d noble metals (Pd or Pt) with small magnetic moments, but large spin-orbit coupling. A layered sample structure of this composition can induce a preferential spin orientation perpendicular to the layers~\cite{Weller1994}. 
Co/Pd and Co/Pt multilayers have received considerable attention as promising candidates for perpendicular magnetic storage and patterned media~\cite{Hellwig2007,Hellwig2009}. While a precise control of the perpendicular magnetic anisotropy has been achieved at the atomic thickness level, a corresponding tuning of the in-plane magnetic anisotropy has not been developed yet~\cite{Ma2015}. 
\par
The effect of quantum well states (QWS) on magnetic phenomena in layered magnetic structures has been studied before ~\cite{Himpsel1998, Dbrowski2014}. QWS may arise when at lest one of the spatial dimensions of the sample has a distinct ratio with the electron wavelength, which is given by the inverse of the electron wave number $k$. QWS are often discussed in the language of electron confinement. This alludes to the physically appealing picture of electrons being trapped, \textit{confined}, between potential barriers. This situation may give rise to a standing wave pattern of the electron density with pronounced spatial oscillations in the direction of confinement~\cite{Fiete2003}. Electron confinement within a nonmagnetic overlayer (NM) on a magnetic substrate is a spin-dependent phenomenon due to the action of the exchange interaction of the substrate. Electron confinement leads to discrete energy levels in the electronic density of states, and these levels shift in energy with a variation of the relevant spatial dimension, \textit{e.g.} layer thickness. The resulting shifting in energy of QWS also leads to a modulation of the electronic density of states near the Fermi level (E$_{F}$). This  induces an oscillatory variation of physical properties such as inter-layer-exchange coupling, magnetic anisotropy and superconductivity as a function of film thickness~\cite{Kunc2015,Parkin1991,Suzuki1998,Shanenko2007}. 

While numerous studies report spin-dependent electron confinement of nearly free \textit{sp} electrons ~\cite{Joly2006,Niizeki2008,Oka2010}, QWS from d electrons are still poorly explored~\cite{Yoshimatsu2011,Saha2014}. The short lifetime and small mean-free path of d-states render them elusive to experimental detection~\cite{Luh2000}. However the confinement of d-electrons is particularly interesting in transition metal films, where the d-electrons largely govern the magnetic properties~\cite{Dbrowski2014}. There is evidence for d-like QWS from spin-resolved photoemission spectroscopy~\cite{Hartmann1993} in Pd(111) films with a thickness of up to 3 atomic layers on ferromagnetic Co. Since these photoemission experiments average laterally over the photon interaction area, they require highly homogeneous films with a constant thickness over the sample area of $\approx$ 0.1 mm. A complementary approach, including spatial resolution on the atomic scale, is provided by scanning tunneling microscopy (STM) and scanning tunneling spectroscopy (STS), where both occupied and unoccupied states can be probed~\cite{Oka2014}.

\par
Two concepts have been proposed to manipulate the magnetic anisotropy (MA) of a ferromagnet (FM) by QWS. One concept exploits QWS formed in the FM itself~\cite{Przybylski2012}, while the other exploits QWS formed in the metallic layer adjacent to the FM film~\cite{Manna2013,Pescia1997}. This metallic layer is often loosely classified as non-magnetic (NM), indicating that the metallic film itself has a vanishing magnetic moment. While numerous theoretical and experimental studies~\cite{Dbrowski2014, Dasa2013, Schfer2007, Cinal2002} confirmed the effect of QWS in ferromagnetic films on the magnetic anisotropy, the role of QWS in the metallic layer on the FM is still poorly explored. 

An oscillatory variation of the MA of Fe and Co films upon a variation of the Cu metallic film thickness has been reported~\cite{Manna2013}. This effect has been attributed to \textit{sp}-QWS formed in the Cu overlayer. The impact on the magnetic anisotropy results from spin-orbit (SO) coupling~\cite{Bruno1989,Skomski2011}, which is affected by the QWS-driven change of the electronic density of states of the adjacent NM. The electronic band structure of the magnetic layer responds by hybridization at the interface or by confinement of electronic states within the potential barriers at the surface or interface. Hence one can anticipate that the evolution of QWS in the NM with increasing layer thickness of a NM material with large spin-orbit coupling leads to stronger changes in the magnetic anisotropy of the adjacent ferromagnetic film, as compared to a NM material with small spin-orbit coupling (Cu). Theoretically, a periodic variation of the magnetic anisotropy energy (MAE) in Pd/Co bi-layers has been predicted~\cite{Cinal1997,Cinal1998}. These calculations suggest that the MAE of ferromagnetic Co is also influenced by spin-polarized QWS formed in the adjacent Pd layers. The Pd/Co system is a promising candidates to elucidate the electronic origin of MAE oscillations due to large SO coupling of Pd. A word of caution on the term non-magnetic is appropriate in this respect. Pd is at the verge of being a FM. Theory predicts~\cite{Cinal1997} a small induced magnetic moment in Pd in close proximity to a FM. This induced FM in Pd affects only the first Pd layers near the interface, and we still regard the majority of the Pd film, which is studied for a thickness of up to 21 ML here, as non-magnetic. Clear MAE oscillations, driven by QWS in Pd, have been predicted for the Pd/Co system in theory~\cite{Cinal2001}. This prediction of large MAE oscillations in the Pd-Co system has been made nearly 15 years ago, but it still awaits its experimental scrutinization. The following study provides experimental evidence in support of this model. 

\par
In this article, we experimentally demonstrate the direct correlation between oscillations of the in-plane magnetic anisotropy of Co films and quantum well states formed in Pd in the Pd/Co/Cu(001) system. Scanning tunneling spectroscopy identifies d-band QWS in Pd. Our MOKE study on Pd/Co/Cu(1,1,13) reveals pronounced Pd thickness dependent variations of the uniaxial magnetic anisotropy of Co, which we ascribe to periodic changes in the density of states at the Fermi level induced by QWS in Pd.  

\section{II. EXPERIMENTAL DETAILS}

Our study reveals the effect of electron confinement in Pd on the magnetic anisotropy of the adjacent ferromagnetic Co film in the Pd/Co/Cu(001) system. This is experimentally challenging, as the expected change of magnetic anisotropy due to Pd QWS is small~\cite{Cinal2001}. To observe this effect two prerequisites need to be fulfilled: (a) in order to form distinct QWS, the Pd films should ideally be grown in a layer-by-layer manner, resulting in the most homogenous film thickness over the MOKE probing area ($0.1$~mm$^2$); (b) the interface between the FM Co film and Pd needs to be atomically sharp. In particular, interface roughness needs to be low to ensure pronounced QWS~\cite{Kloth2014}. The Pd/Co/Cu(001) system (see inset of Figure 1a) fulfills both requirements. This is also due to its low lattice misfit, enabling epitaxial growth~\cite{Meyerheim2007} of face-centered cubic (fcc) Co(001) on Cu(001). The fcc phase of Co has a lattice parameter at room temperature of 3.54 $\AA$~\cite{Cerda1993}.   
\par
The experiments were carried out in two ultra-high vacuum (UHV) chambers (with base pressure below 2x10$^{-10}$ mbar). One dedicated to low temperature scanning tunneling microscope/spectroscopy (LT-STM/STS), the other dedicated to magneto-optical Kerr-effect (MOKE) experiments. A Cu(001) and a vicinal Cu(1,1,13) substrate [6.2$^{0}$ off to (001) surface] were
prepared by cycles of 1 keV Ar$^{+}$ ion bombardment (25-30 min) and subsequent
annealing at 900 K (40 min). The Co films were grown at 170 K by molecular beam epitaxy(MBE) at the rate of 0.5 monolayer (ML)/min. One ML is defined as the surface atomic density of Cu(001). After growth, the films were annealed at 300K, in order to reduce surface roughness. Pd was deposited at 300 K with a wedge-like spatial thickness variation (see inset of Fig.~1(a)). Thus, the Pd thickness increases along the [-110] direction of the Cu substrate by 3 ML per mm. The chemical cleanliness, the surface structure, and the surface morphology were investigated \textit{in-situ} by Auger electron spectroscopy (AES), low energy electron diffraction (LEED), reflection high energy electron diffraction (RHEED) and scanning tunneling microscopy (STM). All STM/STS measurements were done at 4.7 K. STM tips were prepared from polycrystalline tungsten (W) wire, chemically etched and flashed to 2200 K. STS data were obtained using a lock-in technique to record the differential tunneling conductance (dI/dV) by adding an AC (frequency = 3.6 kHz) modulation voltage V$_{rms}$= 10 mV to the bias voltage, while ramping the applied bias V.    
\par
Magnetic properties were probed by \textit{in-situ} longitudinal MOKE. Our static MOKE system has been described previously~\cite{Przybylski2012}. A laser diode provides light of wavelength 670 nm with a beam diameter $<$ 0.2 mm at the sample surface.  We use a fixed incidence angle= 21$^{0}$ with respect to the sample normal. The s-polarized light is reflected from the sample, passes through a quarter-wave plate, an analyzer and is detected by a photodiode. The photodiode signal (Kerr signal) corresponds to the reflected beam intensity, and it depends on the sample magnetization. This signal is recorded as a function of the external magnetic field, which was applied in the film plane and along the optical plane, to obtain the hysteresis loop of the Kerr signal. The sample was placed in a specific MOKE manipulator, which allows to rotate the sample azimuthally by $\pm$360$^{0}$. The laser beam diameter at the sample surface is of order 0.2~mm. This laser footprint covers a thickness range of the Pd wedge (slope$~$ 3 ML/mm) sample of $\pm 0.6$~ML. The MOKE measurements were performed in a temperature range from 300 to 5 K.

\section{III. RESULTS AND DISCUSSION}
Figure~1(a) shows a constant current STM image of the Cu(001) substrate. A single atomic step separates the lower terrace (dark, left) from the upper terrace (bright, right). The inset shows a sketch of the Pd wedge. Figure 1(b) presents a constant current STM image of Cu(001) with a Co coverage of 10 ML. The exposure of only three height levels (9, 10 and 11~ML), with negligible contributions from the layer below (8 ML, dark), reveals decent layer-by-layer growth of Co. Figures 1(c,d) show the morphology for the Pd/10 ML Co/Cu(001) with an average Pd thickness of 2.8 ML (c) and 8.5  ML (d). Note, that only two adjacent Pd layer thicknesses are seen, indicative of good layer-by-layer growth of Pd on Co/Cu(001). We perform STS on these surfaces for different thicknesses of Pd. STS shows discrete peaks, which are ascribed to QWSs in Pd, see Fig.~2. 

\begin{figure}
\begin{center}
\includegraphics[width=0.50\textwidth]{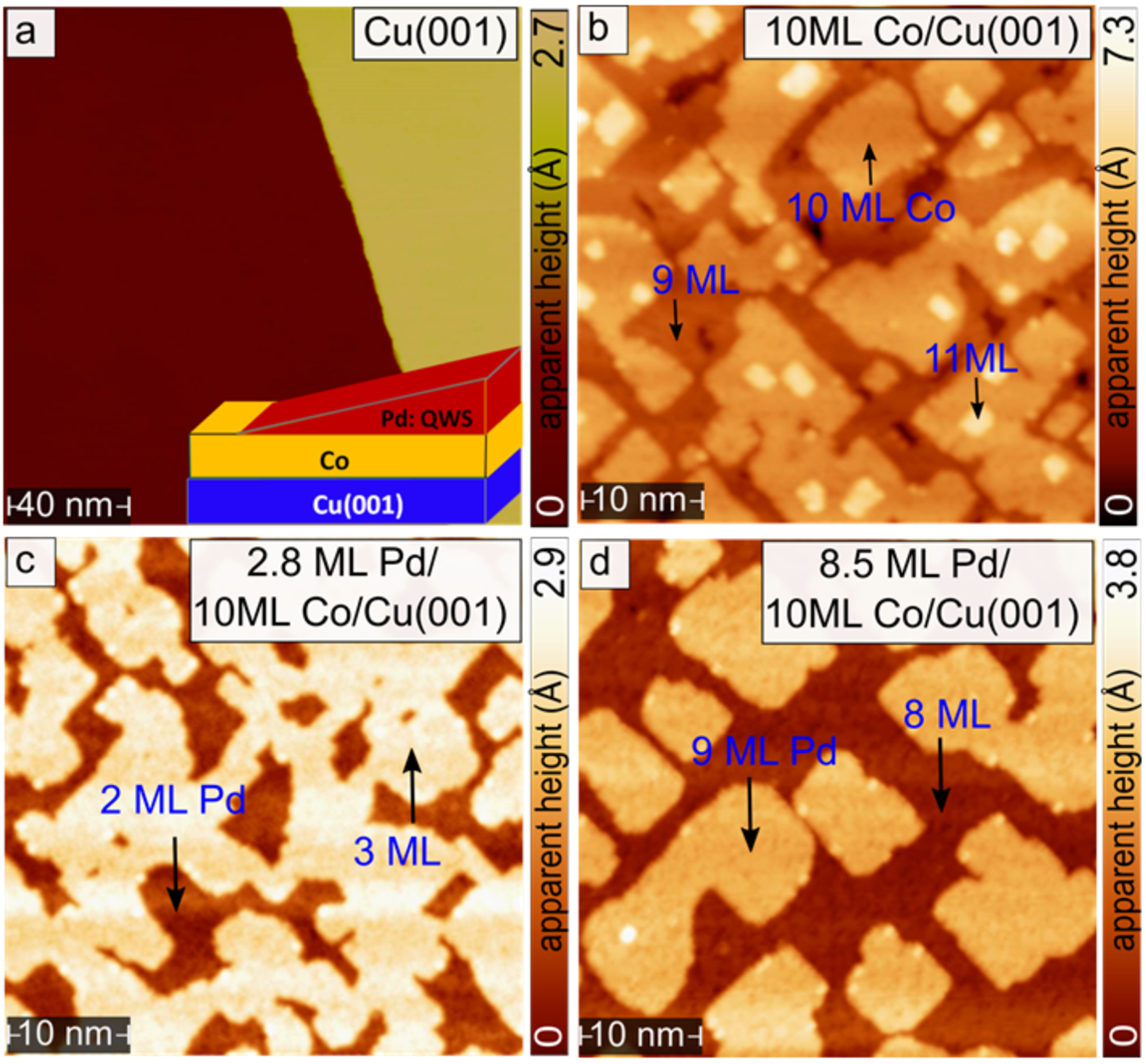}
\caption{(Color online) STM constant current images of different surfaces. (a) The Cu(001) substrate shows two terraces, separated by a mono-atomic step (200 x 200~nm$^{2}$, I = 0.1~nA, V = 1~V). The inset presents a sketch of the Pd(N)/Co/Cu(001) system. (b) 10 ML Co on Cu(001) (50 x 50~nm$^{2}$, I = 0.5~nA, V = 0.3~V), (c) 2.8 ML  Pd/10 ML Co/Cu(001) (50 x 50~nm$^{2}$, I = 1.3~nA, V = $-1.2$~V). (d) 8.5 ML Pd/10 ML Co/Cu(001) (50 x 50~nm$^{2}$, I = 1.5~nA, V = 0.2~V).}\label{Fig1}
\end{center}
\end{figure}

\par
Figure 2(a) displays a series of differential conductance (dI/dV) spectra measured on different Pd(N) thicknesses on (N) Pd/10ML Co/Cu(100). Well defined sharp peaks are visible, see black arrows, within the bias range from $-1$ to $+1$~V around the Fermi energy. The peaks are ascribed to QWS in Pd. The number N on the right-hand side of each spectrum represent the Pd thickness in ML. The number of peaks increases with increasing Pd film thickness. The energy position of the QWS changes with respect to the Fermi level with increasing film thickness. QWS cross the Fermi energy with a thickness periodicity of 6 ML Pd. Thus, the formation of QWS modulates the density of states near E$_{F}$. 

\begin{figure}
\begin{center}
\includegraphics[width=0.50\textwidth]{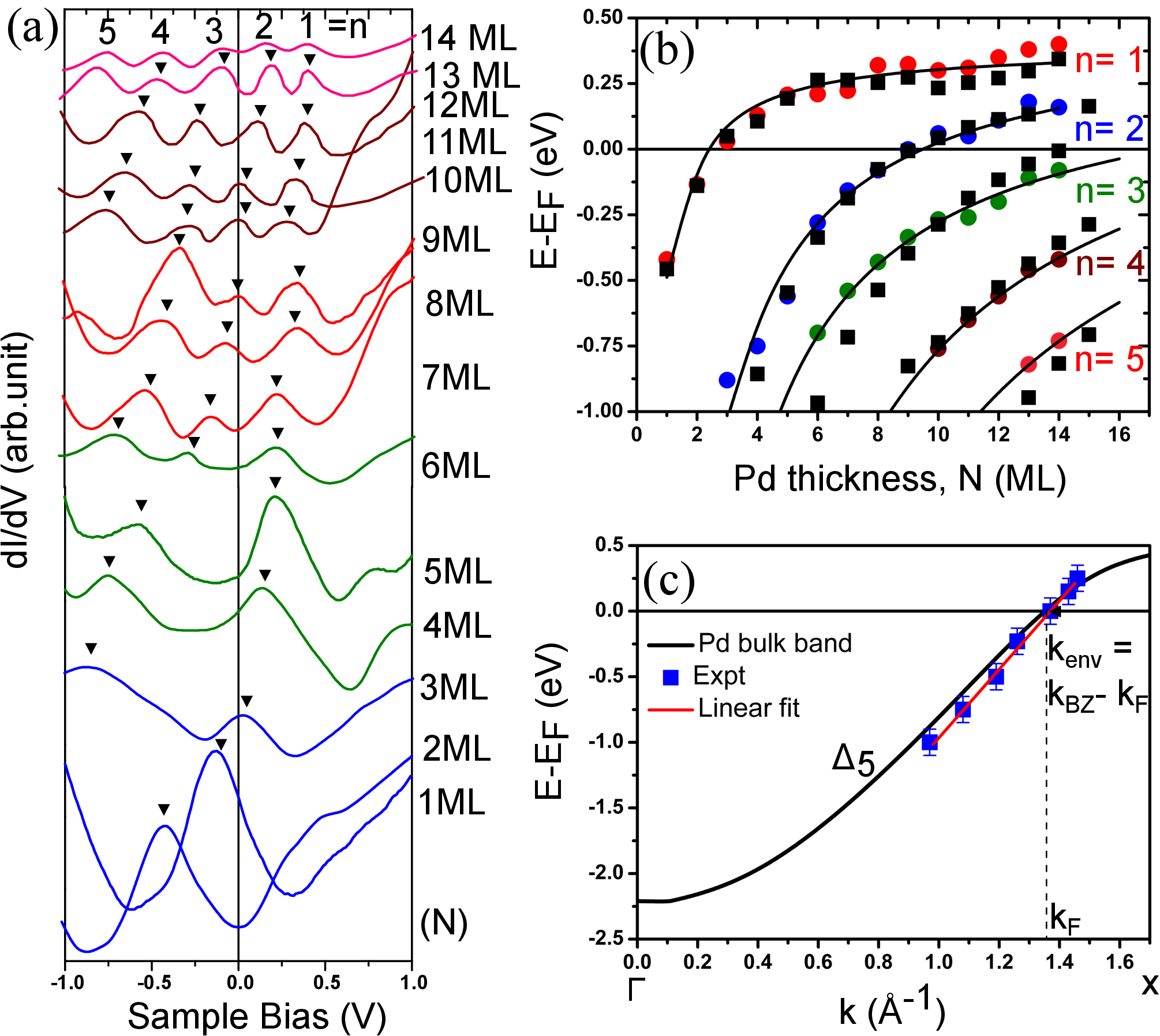}
\caption{(Color online) (a) Series of dI/dV spectra measured on different thickness of (N) Pd/10ML Co/Cu(001). Prior to the spectroscopy, the tunneling gap was stabilized at 3 nA and 1 V. Spectra are vertically offset for clarity. The labels (N) besides the graph show the thickness of the Pd film. (b) Quantum well peak energies (taken from dI/dV spectra) as a function of Pd thickness (colored solid circles) and fitted energies of QWS obtained from the phase accumulation model (black solid squares). The number n = 1 to 5 represents an index of QWS. (c) Experimentally determined (using equation 2) dispersion k(E) together with the Pd d-band ($\Delta$$_{5}$, as  taken from Mirbt.et.al.~\cite{Mirbt1996})(solid line) along the ($\Gamma$-X) direction in the 1st BZ. The red line represents a linear fit of k(E).}\label{Fig2}
\end{center}
\end{figure}

\par
To discuss the formation of QWS in the Pd films, we consider the energy bands of bulk Pd(001) in the direction($\Gamma$-X), perpendicular to the interface, along the film normal~\cite{Mirbt1996}. Bulk Pd(100) has fairly dispersive $\Delta$$_{5}$ bands (large degree of d-character), which originate mostly from d$_{zx}$ and d$_{zy}$ states~\cite{Saha2014}, in the energy range, where QWS are visible in the tunneling spectra. Pd d states are confined to a quantum well in the Pd film, where the Pd film thickness is the decisive quantization parameter. 

To gain a quantitative understanding, we fitted the QWS energy positions in the energy--thickness plane using the so-called phase accumulation model (PAM)~\cite{Kloth2014,Uchihashi2008,Qiu2002,Shikin2002,Ortega1993,Smith1985,Echenique1975}, where electrons inside Pd are confined in a potential well of width N$_{Pd}$. This corresponds to the quantum mechanics example of a particle in a box. The quantization condition for an electron state in such a potential box is given by
 
\begin{multline}
2n\pi =2k({E}_n)Nd + {\Phi}_s(E_{n}) + {\Phi}_i(E_{n})
\end{multline}

where n is a quantum number, k(E) is the Pd band dispersion in the direction ($\Gamma$-X) along the film thickness, N is the number of atomic layers of the film, d=1.89 $\AA$ is the lattice spacing along the $\Gamma$-X direction, k is the wave vector perpendicular to the film plane, and $\Phi$$_{s}$ and $\Phi$$_{i}$ are phase shifts upon reflections at the surface and the interface, respectively~\cite{Shikin2002}. We determined the appropriate phase value~\cite{Smith1994} from the boundary condition of the Pd bulk band structure, and this gives us the possible QW energy for each thickness of Pd using Eq.1. 

Figure 2(b) displays the discrete QW energy extracted from the fitting (black square). The fitted values correspond convincingly to the experimental QW energies, taken from the individual peaks of the differential conductance spectra (solid circles). We also fit the QWS in the E-d plane [fig.2(b)] by assuming a linear dependence of the total phase shift on the energy; i.e., $\Phi$$_{s}$ + $\Phi$$_{i}$ = $\pi$(a$^{*}$E+b). The fitting result (black solid line) is in excellent agreement with the experimental results. STM was crucial to extract the local Pd thickness accurately. The thickness determination error is estimated as $\pm$1 ML. 
\par
An important application of quantum well spectroscopy is the band structure determination for both occupied and empty state. The basic idea is that k$_{\bot}$ is quantized in a film as a result of electron confinement. Measurements of QWS peak positions for many different film thicknesses should permit a unique solution of E(k$_{\bot}$). Because the phase shift depends on electron energy, the wave vector k(E) at a given energy can be derived from the oscillation period of the QWS at that energy. Assuming that for two Pd thicknesses N$_{1}$ and N$_{2}$ one can find the respective spectroscopic peaks E$_{n1}$ and E$_{n2}$ such that E$_{n1}$=E$_{n2}$=E, with index n$_{1}$ and n$_{2}$, the corresponding bulk band dispersion along $\Gamma$-X can then be obtained~\cite{Saha2014} as 

\begin{multline}
k(E) = \pi({n}_2 - {n}_1)/[({N}_2 - {N}_1)d] + C.
\end{multline}

\noindent The correction term C= (${\Phi}$$_{2}$-${\Phi}$$_{1}$)/2(N$_{2}$-N$_{1}$)d comes from the resulting phase shifts at both the Pd-vacuum and Pd-Cu interfaces due to finite height of the potential barriers forming the well. We determined the band dispersion k(E) for Pd QWS using Eq.2, as shown in Fig. 2(c). For a direct
comparison, the dispersion curve is plotted with a bulk Pd d-electronic ($\Delta$$_{5}$) band taken from Mirbt. et. al.~\cite{Mirbt1996}. The experimentally derived dispersion is consistent with the theoretical one. The energy band near E$_{F}$ exhibits an approximately linear relation between E and k. Since this energy band comes
from band folding from the second Brillouin zone (BZ), we fit the band using E = $-\frac{\hbar^{2}}{2m_{e}}\cdot(2k_{BZ}-k_{F})(k-k_{F})$. The fitting (shown by the red line in Fig.~2c) yields a Fermi wave vector of k$_{F}$=0.238\AA$^{-1}$ and k$_{F}$$^{'}$= 2k$_{BZ}$-k$_{F}$=1.735 \AA$^{-1}$. The value of k$_{F}$ leads to a Fermi level crossing of the QWS at every $L$(period)= $\pi$/k$_{F}$d(ML)= 6.25 ML, in agreement with the previous calculation~\cite{Cinal2001}. This result is consistent with the period derived from the Fermi wave vector extracted from the bulk band of Pd~\cite{Niklasson1997}. The period $L$ is directly related to the wave vector of the confined electron state spanning the crossing point of the electronic band at a given energy and the nearest high symmetry point of the Brillouin zone (BZ)~\cite{Smith1985}. The wave vector corresponding to a crossing of the Fermi level by the $\Delta$$_{5}$ band is comparable. From the above analysis, we can conclude that the electrons from d-band with $\Delta$$_{5}$ symmetry are mainly confined to form QWS in the Pd films. We have thereby established the origin of the QWS in the Pd film and explained why they change with Pd thickness. 
\begin{figure}
\begin{center}
\includegraphics[width=0.50\textwidth]{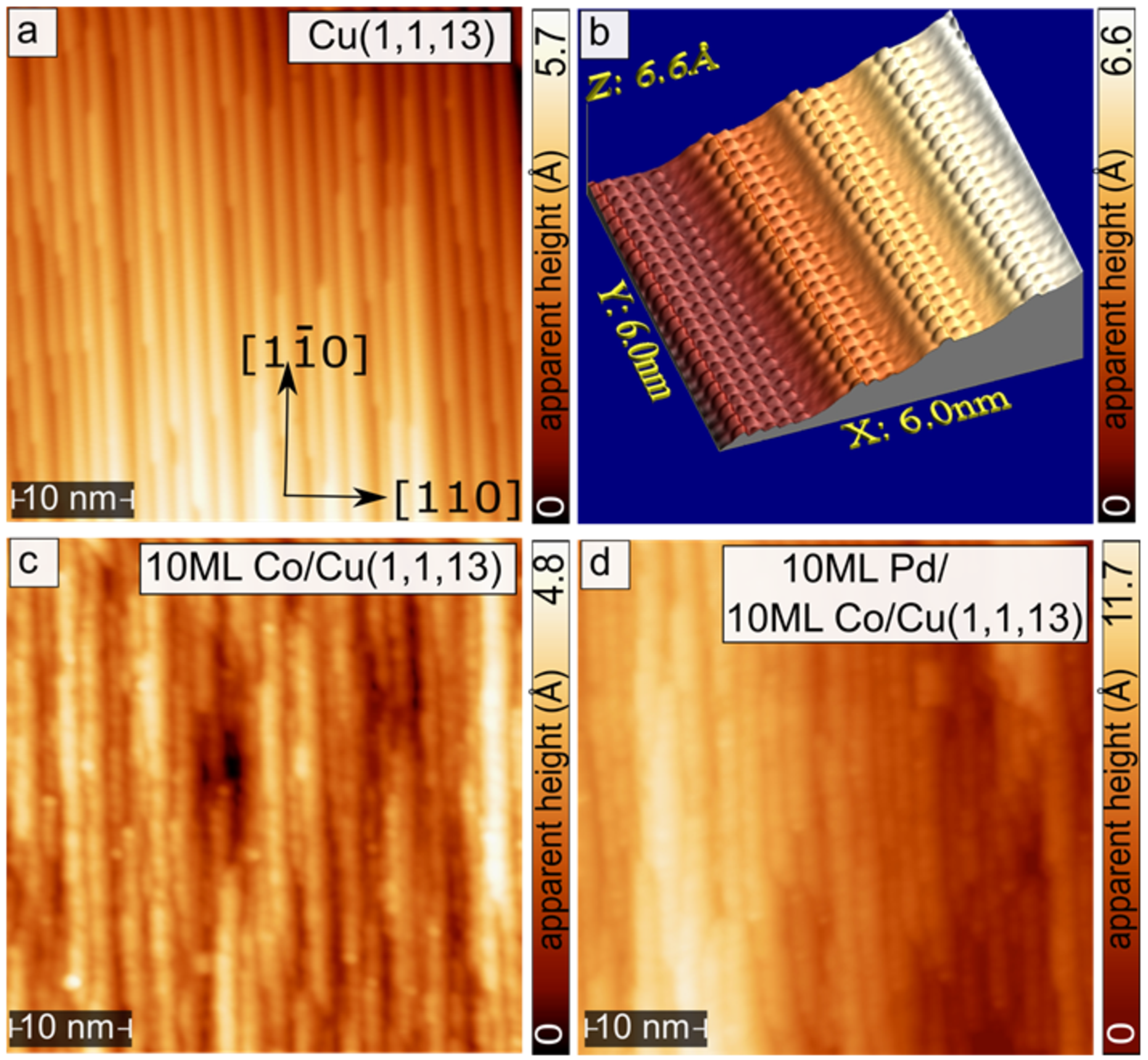}
\caption{(Color online) Constant current STM image (50 x 50 nm$^2$, I = 1~nA, V = 0.5~V) of (a) Cu(1,1,13), (b) atomically resolved STM image of Cu(1,1,13) (6 x 6 nm$^2$, I = 50~nA, V = 21~mV), (c) constant current image of 10 ML Co/Cu(1,1,13) (50 x 50 nm$^2$, I = 1~nA, V = 0.2~V) and (d) 10 ML Pd grown on 10 ML Co/Cu(1,1,13) (50 x 50 nm$^2$, I = 0.8~nA, V = 0.35~V).}\label{Fig3}
\end{center}
\end{figure}

\begin{figure}
\begin{center}
\includegraphics[width=0.50\textwidth]{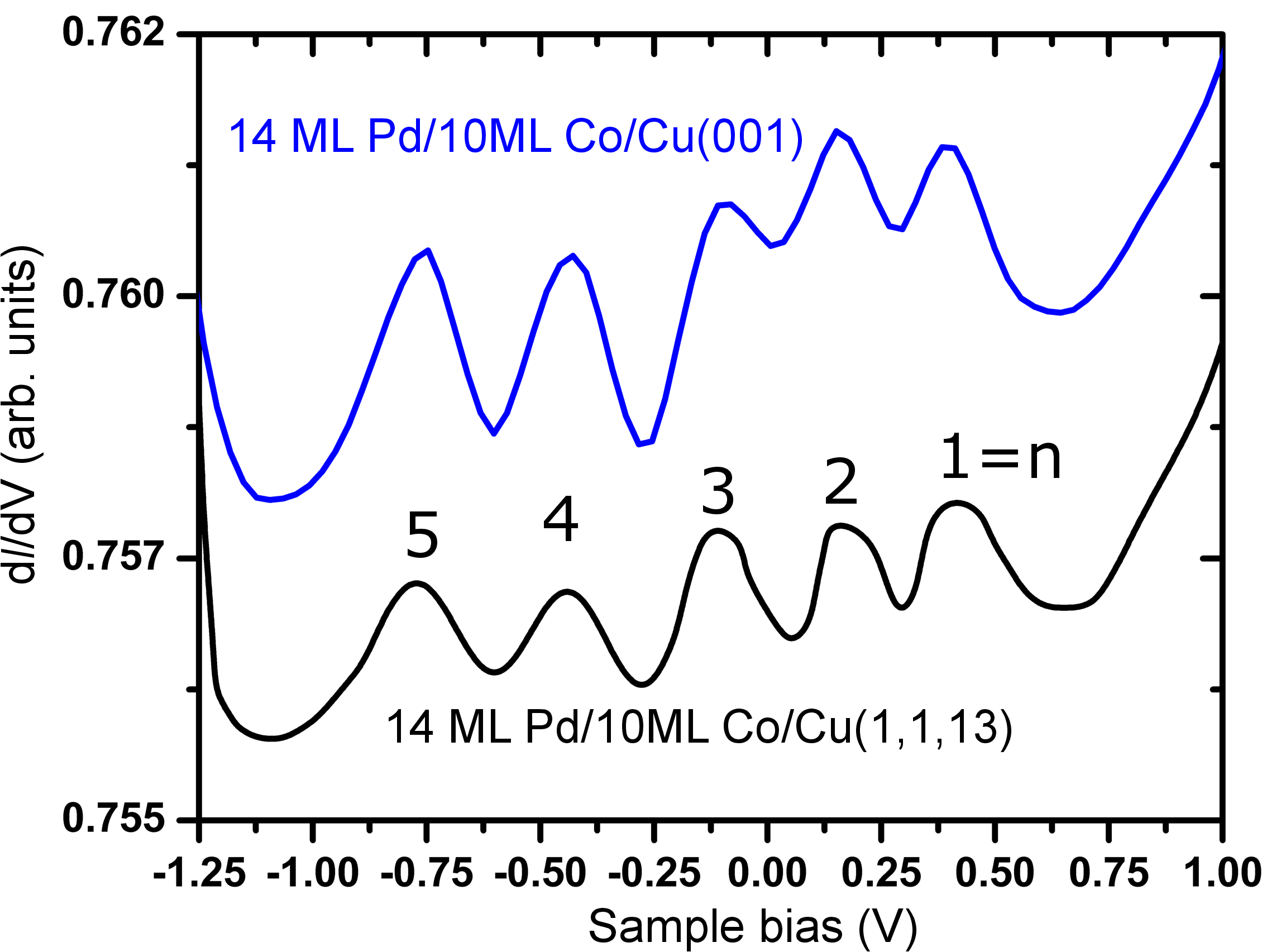}
\caption{(Color online) Differential conductance (dI/dV) spectra measured on 14 ML Pd/10 ML Co/Cu(001) (top curve) and on 10ML Co/Cu(1,1,13) (bottom curve). The number n= 1 to 5 represents an index of QWS, as also introduced in Fig.~2(b) above. The spectra are shifted vertically for clarity.}\label{Fig4}
\end{center}
\end{figure}

\par
In order to study the effect of QWS in Pd on the magnetic anisotropy of the Pd/Co/Cu system, we performed MOKE measurements on Pd/Co films, which were grown on a Cu(1,1,13) vicinal surface. This substrate can be viewed as a stepped Cu(001) surface, where the individual terraces are 1.6~nm wide, separated by monoatomic steps. This surface morphology gives rise to an uniaxial in-plane magnetic anisotropy of Co grown on this surface. The additional uniaxial in-plane anisotropy is ascribed to the symmetry reduction by the steps~\cite{Berger1992}. In our experiments we follow how QWS in Pd modify the uniaxial in-plane magnetic anisotropy of Co. The determination of magnetic anisotropy from hysteresis loops is non-trivial, and it is based on assumptions concerning the magnetization reversal mechanism~\cite{Weber1997, Ouazi2012}. This limitation is overcome in this system with a superposition of twofold and fourfold anisotropies. Co films grown on Cu(1,1, 13) with a well defined preferential step arrangement belong to this class of systems~\cite{Berger1992}. 
\par
To make this approach work, it is necessary that the Co films grown on a stepped surface replicate the step morphology of the surface. The Cu(1,1,13) single crystal used in this study had a preferential step edge direction along [1-10] with an average terrace width of 6.5 atomic distances. The Cu(1,1,13) surface and the morphology of Co and Pd on top has been investigated by STM. Constant current topographic images of the Cu(1,1,13) crystal are shown in Fig. 3(a and b). The (001) terraces are separated by monoatomic steps, where the steps run predominantly along [110]. The average terrace width measured from atomically resolved STM image is $\approx$ 1.56~nm. This is very close to the expected value for this surface, and it agrees very well with the LEED measurements (see the supporting information). Cu(1,1,13) consists dominantly of (001) terraces. We observe the same growth behavior for Co on Cu(1,1,13) and on Cu(001). The STM image (see Fig.~3(c)) for 10 ML Co grown on Cu(1,1,13) identifies a  layer-by-layer growth of Co, which reproduces the step density of the substrate favorably. A close inspection reveals that the top Co terraces tend to be slightly broader ($\approx$ 2~nm) in comparison with the substrate terraces. Still, well-aligned monoatomic steps are clearly visible. 

Previous STM studies on Co films grown on vicinal Cu(001) surfaces~\cite{Giesen1995, ChauminMidoir2002} have shown that at lower coverages of Co, (up to $\approx$ 5 ML), the surface is rugged and there is no clear preferential orientation of islands. Upon increasing the Co thickness, the islands coalesce and form straight steps elongated parallel to the step edges of the substrate. It was proposed that the initial roughening of the step structure is caused by the minimization of the strain energy~\cite{Giesen1995}.

Layer-by-layer growth of Pd on 10 ML Co/Cu(1,1,13) is observed, as shown by STM (Fig. 3(d)). Since the MA of ultrathin films is inherently connected to the structure and morphology of the films, it is essential that the idealized film growth picture, leading to a direct replica between substrate and film, is experimentally approximated in the best possible manner. This is given here.
\par
One might argue that the QWS measured in STS on the Pd film grown on Co/Cu(100) would not be identical to the QWS measured in Pd/Co/Cu(1,1,13). In order to check this, we performed STS measurements for distinct  Pd thickness on Co/Cu(1,1,13). Figure 4 reveals that for 14 ML of Pd, the QWS energies and the overall spectral shape are comparable and show the same general features. We conclude that the electronic structure of the Pd film is fully comparable for growth on the flat and the vicinal Cu substrate.

\begin{figure}
\begin{center}
\includegraphics[width=0.50\textwidth]{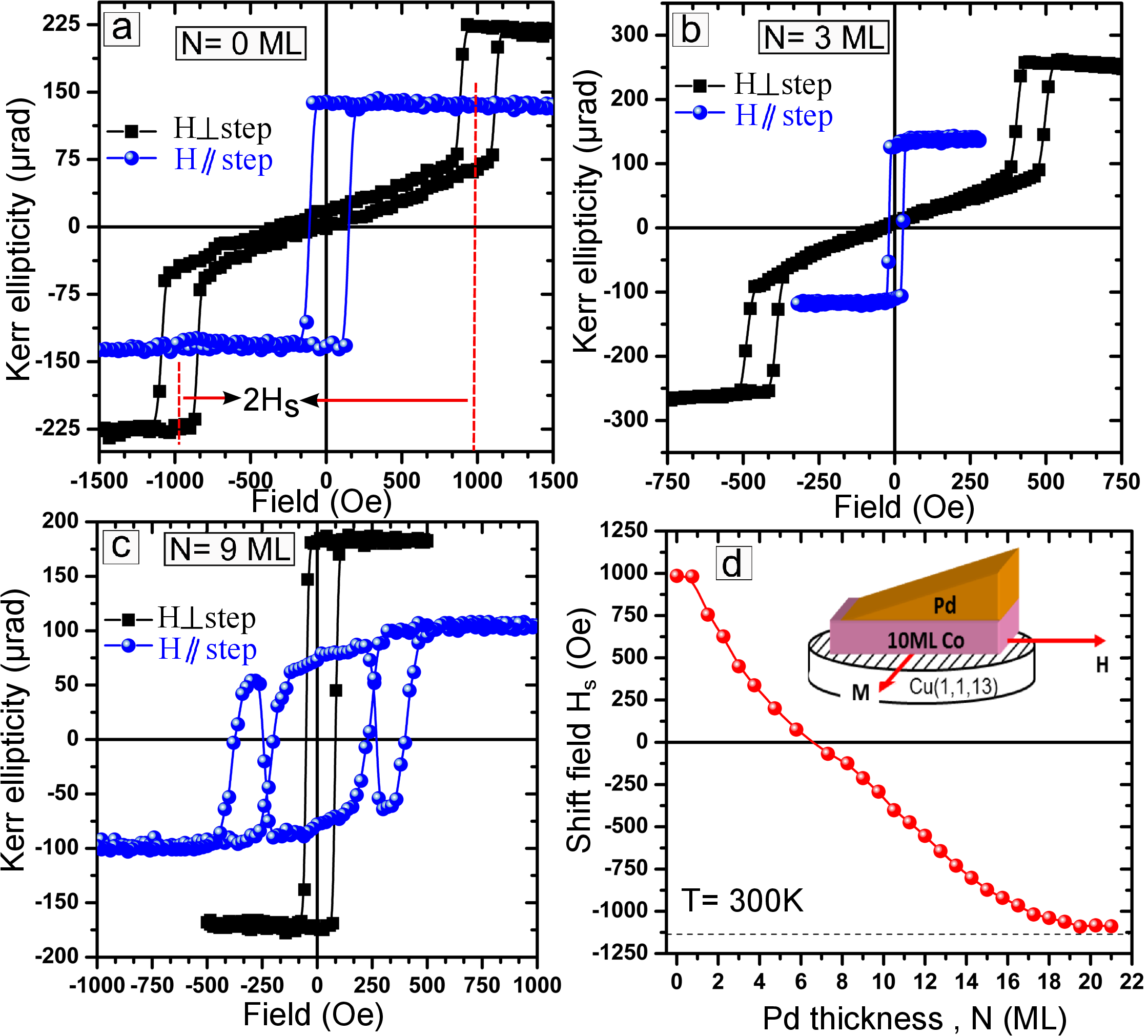}
\caption{(Color online) (a-c) Hysteresis loops measured at T = 300 K for different Pd thickness (a) 0 ML, (b) 3 ML and (c) 9 ML, respectively. Measurements were performed with the magnetic field (H) applied perpendicular (black) and parallel (blue) to the step edge orientation. (d) Plot of the shift field H$_{s}$ measured at T = 300 K for different Pd thickness in Pd(N)/10 ML Co/Cu(1,1,13). Positive and negative H$_{s}$ indicate that the easy magnetization axis is oriented along or perpendicular to the step edge, respectively. The inset shows a  schematic of the Pd wedge on Co/Cu(1,1,13).}\label{Fig5}
\end{center}
\end{figure}

\par
Magnetic hysteresis loops were recorded \textit{in-situ} for Pd in the thickness range between N = 0 to 21 ML grown on 10 ML thick Co films on Cu(1,1,13). Figure 5(a) shows the resulting hysteresis loops. The magnetic field is applied either parallel (along[1-10]) or perpendicular (along [110]) to the step edges of the Cu substrate. The easy magnetization axis runs parallel to the step edge, as shown by the rectangular hysteresis loop (blue curve) for zero Pd coverage in Fig.~5(a). The curve measured perpendicular to the step edges is more complicated (black curve). It shows two single loops shifted against each other by the shift field $H_{s}$. This difference of the magnetic response for magnetization along the [1-10] and the [110] directions is the signature of the uniaxial magnetic anisotropy of this system, where the uniaxial anisotropy is given  the preferred step orientation of the vicinal surface ~\cite{Bruno1988,Weber1996}. For fcc Co on Cu(001) both directions are structurally and magnetically equivalent, and an easy magnetization direction along any [110] direction is observed~\cite{Schneider1990}.
\par
A significant difference in saturation Kerr signal of 10 ML Co for measurements parallel (smaller) and perpendicular (larger) to the step edge (Fig. 5 (a) is observed. This has been observed before for Co/Cu(1,1,13)~\cite{Bauer2011b}, and the larger MOKE signal for measurements perpendicular to the step edge is ascribed to an effective polar MOKE signal component. It can be ascribed to the macroscopically inclined surface of the stepped sample in the direction perpendicular to the step edge. A corresponding component is absent for measurements along the step edge. Since the polar Kerr effect is much stronger than the longitudinal one (roughly one order of magnitude), even a small normal component of the magnetization can give a significant contribution to the total Kerr signal. The normal component of the magnetization can have an influence on the split hysteresis loops measured by probing the magnetization along the steps (i.e. along the intermediate axis). When probing the magnetization along the steps, only the Kerr signal at low field will be influenced by the normal component of the magnetization (see in Fig. 5c). This is also related to the competition between shape, in-plane, and perpendicular magnetic anisotropy which tilts the magnetization off from the sample plane~\cite{Bauer2011a}. 
\begin{figure}
\begin{center}
\includegraphics[width=0.50\textwidth]{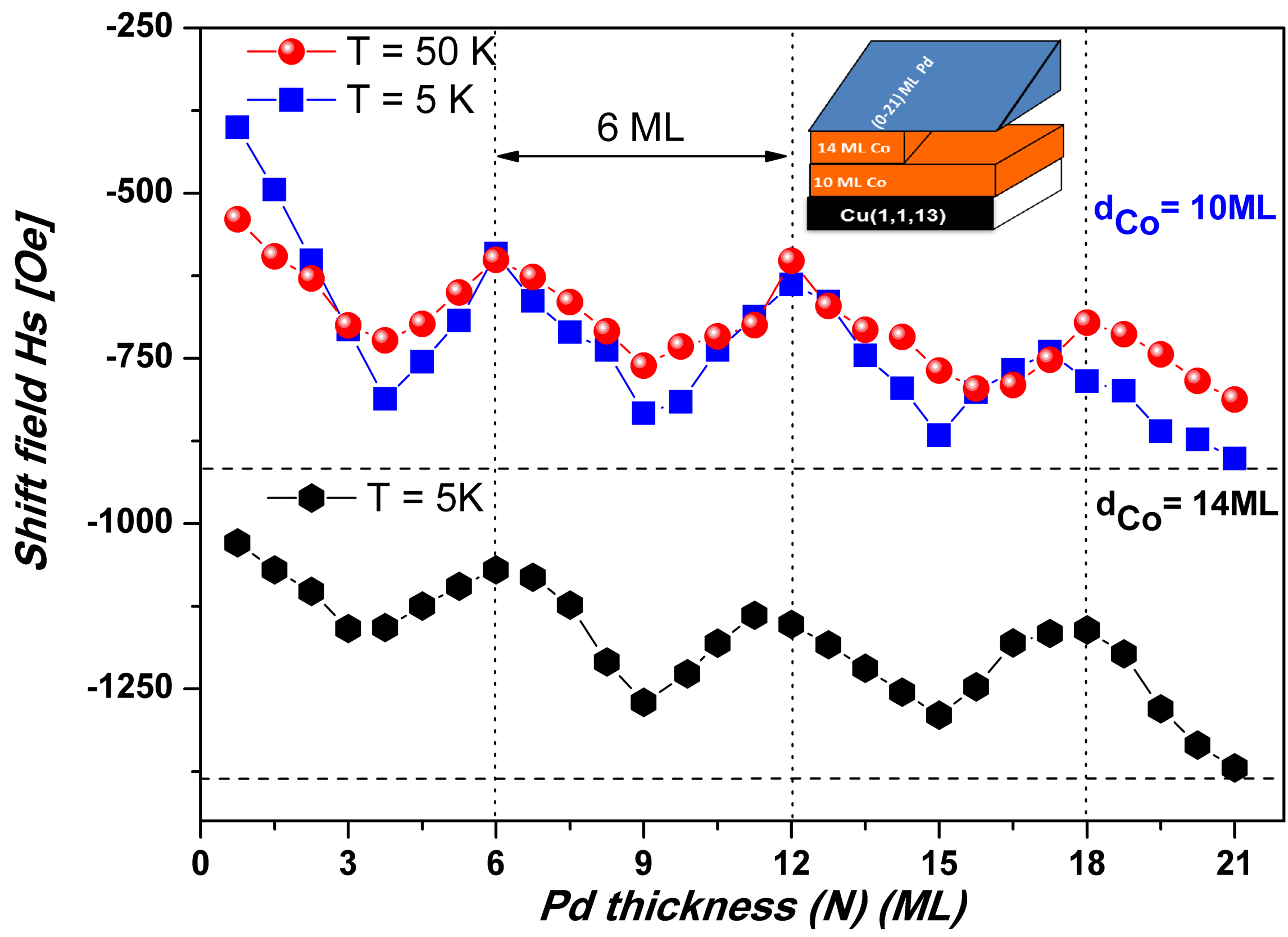}
\caption{(Color online) Upper panel. Switching filed H$_{s}$ in dependence on the Pd thickness for Pd(N)/10ML Co/Cu(1,1,13), measured at T = 50 K (red dots) and T = 5 K (blue dots). Lower panel. Shift-field H$_{s}$ measured at 5 K, in dependence on the Pd thickness for 14 ML Co/Cu(1,1,13) (black dots). The inset represents a sketch of the double wedge sample, which provides both 10 and 14 ML Co.}\label{Fig6}
\end{center}
\end{figure}
\par
For the stepped surfaces studied here, we call the [1-10] direction the easy magnetization axis, and the [110] direction is the intermediate axis. This reflects that the latter combines the easy character of the fourfold cubic anisotropy with the hard character induced by the step-orientation and the resulting uniaxial anisotropy. A composite hysteresis loop results for magnetization along the intermediate axis. The shift field H$_{s}$ is defined as the field difference between zero field and the center of one of the shifted loops. Positive and negative H$_{s}$ indicate that the easy magnetization axis is oriented along or perpendicular to the steps. In a simplified model of magnetization reversal by coherent single-domain switching, the shift field H$_{s}$ is proportional to the in-plane uniaxial magnetic anisotropy K$_{u}$ ~\cite{Weber1997}. This relation has been shown to be a reliable approximation, particularly for small variations of K$_{u}$~\cite{Pescia1997,Cinal2006,Li2009}. 
\par
A detailed description of the magnetization reversal~\cite{Weber1995a}, however, is more complex. It demands taking into account the domain wall energy and avoiding the separation of the in-plane fourfold and in-plane twofold anisotropies~\cite{Hyman1998, Oepen2000}. Still, the value of H$_{s}$ is a measure of the in-plane uniaxial magnetic anisotropy induced by the steps~\cite{Kawakami1996,Li2009,Bauer2011a}. Its measurement  allows us to detect with high sensitivity and accuracy the possible impact of Pd QWS on the magnetic anisotropy of this uniaxial system. 

Figure~5(d) presents a plot of H$_{s}$ in dependence on the Pd thickness for Pd(N)/Co/Cu(1,1,13). Intriguingly, the shift field H$_{s}$ changes pronouncedly with Pd film thickness. This result reflects a Pd-driven variation of the step-induced anisotropy of Co. The H$_{s}$ value decreases up to 6 ML Pd coverage, and it stays constant  above 18 ML Pd. For a Pd coverage above 6 ML, the characteristic easy axis of the 10 ML Co films switches from along the steps to perpendicular to steps, and this is reflected in the H$_{s}$ plot as zero transition with a corresponding change of sign. The split hysteresis loop of the perpendicular direction   
 is replaced by the rectangular loop of an easy magnetization axis, where the parallel($||$)-direction looses its easy character, see Fig.~5(c).

How can the Pd coverage modify the magnetic anisotropy of the Co film? At first sight one might speculate that the electronic structure and the structurally relaxation at step edges gets modified upon Pd coverage~\cite{Brovko2014}. It is plausible to assume that a strong change of the magnetic anisotropy at the step site results, changing the uniaxial anisotropy. However, the observed periodic change of the magnetic anisotropy remains mysterious in this picture. The complex hysteresis loops observed (see fig.5c) for higher Pd coverage (H $||$ to the step) can be explained as the superposition of the split loops with the reverse hysteresis loops (at low field)~\cite{Bauer2011a}. Comparable effects of a change of anisotropy have been previously observed for different adsorbates such as Cu, Ag and O~\cite{Weber1995b, Engel1993}. They were strong enough to change the easy magnetization axis by 90$^{0}$ in-plane. This scenario would correspond to an interfacial effect. 

An interfacial effect would also result from periodic electronic changes near the Pd/Co interface, which modulate the anisotropy of the Co through spin-orbit coupling and exchange interaction between Co and Pd. But according to the calculation by Cinal ~\cite{Cinal2001}, we are not dealing with an interface effect. Instead, the magnetic anisotropy oscillations originate almost entirely within the Pd overlayer, and they are transferred through the Co-Pd interface into the Co system. Therefore, an important question is whether the period of magnetic anisotropy oscillations depends on the thickness of the Co film or not. To tackle this experimentally we repeated the measurement for two samples with different Co film thickness (10 and 14 ML), where two different strengths of uniaxial anisotropy are expected. 

QWS related effects are usually strongest and best observed at low temperature ~\cite{Manna2013}. Consequently, we investigate the dependence of the switching field H$_{s}$ on the Pd thickness(N) for Pd(N)/10ML Co/Cu(1,1,13) at low temperature. Figure 6 shows the stunning result. The switching field H$_{s}$ shows periodic maxima (N=3, 19, 15 ML) and minima(N=6, 12, 18 ML) as a function of Pd thickness. The period of this oscillation is $\approx$6 ML. The same periodicity is observed for measurements at 50 and 5 K. The perodicity also remains constant for a different Co thickness of 14 ML, as revealed by the top and bottom panel of Fig. 6, respectively. The shift field itself depends on the Co thickness. This reflects that H$_{s}$ itself is modified by QWS formed in Co. But the periodicity of the in-plane magnetic anisotropy variation remains constant at 6 ML Pd. Interestingly, the amplitude of the magnetic anisotropy oscillation are very similar for both Co film thicknesses. All these results point at the importance of the Pd layer and its thickness for the observed variation of the in-plane anisotropy, and this diminishes the importance of interface effects.

\par
The oscillation amplitude($\approx$150 Oe) is almost three times larger than the anisotropy oscillation caused by QWS in Cu in the Cu/Co/Cu(001) system~\cite{Manna2013}. This large amplitude of magnetic anisotropy oscillation is comparable to the previously observed effect of QWS on the magnetic anisotropy of Co films grown on vicinal Cu(001)~\cite{Bauer2011b}. Qualitatively, QWS formed in the Pd overlayer offer a natural  driving force for modifying the MA of the Pd/Co/Cu system. QWS modulate the density of states near the Fermi level. The magnetic anisotropy is largely determined by states in proximity to the Fermi energy~\cite{Wu1998}, and hence a modification of the magnetic anisotropy is expected. A more thorough analysis of the origin of the change of MA is given below. 

Both, the changes of magnetic anisotropy with increasing thickness of Pd and the STS studies on QWS in Pd identify the same characteristic periodicity of 6 ML Pd. This clearly suggests that the QWS in Pd are a key aspect for the oscillatory magnetic anisotropy of Co. From the STS analysis we showed that the Pd QWS have predominantly d character ($\Delta$$_5$ a-like). In principle, theses QWS should therefore affect the anisotropy significantly, as the magnetic anisotropy of transition metals such as Co is mainly governed by d-bands. 
\par
Cinal has studied the impact of a Pd coverage on the magnetic anisotropy of Co~\cite{Cinal2001}. The author suggests that the magnetic anisotropy oscillations in the Pd/Co system are induced by pairs of spin-polarized QWS in Pd, where each pair is degenerate at the $\Gamma$ point of the two-dimensional surface Brillouin zone. The majority spin QWS, forming the pairs, were derived from bulk Pd electronic states, whose three-dimensional wave-vector were quantized in the direction perpendicular to the surface~\cite{Cinal2001}. An experimental support for this comes from our STS studies which identify corresponding QWSs. Minority spin QWS in Pd hybridize with Co d states and form a resonance. As a result a periodic change of the electronic structure of the ferromagnetic Co leads to the oscillation of the magnetic anisotropy, which is mediated by the large spin-orbit coupling of Pd. Note that the magnetic anisotropy is a quadratic form of spin-orbit coupling. The large amplitude of MA oscillations in Pd/Co as compared to the Cu/Co system is a cnsequence of the larger spin-orbit interaction in Pd as compared to Cu. Theory also predicts that the anisotropy oscillation period is governed by the extremal dimension of the bulk Pd Fermi surface. Our experimental results corroborate this prediction. The observed damping of MA oscillations at room temperature speaks in favor of an effect driven by electronic states near the Fermi level. Thus, our experiments support fully the theoretical prediction of the dominant role of QWS in Pd for the magnetic anisotropy of Co in the Pd/Co/Cu system.

\section{Conclusions}
Our combined STM and MOKE studies reveal the effect of electron confinement and QWS in Pd on the  uniaxial magnetic anisotropy of ferromagnetic Co in Pd/Co/Cu. Pseudomorphic growth, well-defined interfaces, and STS with single layer thickness resolution allow us to probe occupied and unoccupied QWS and extract an accurate dispersion of the Pd electronic d-band. The magnetic anisotropy is found to oscillate as a function of Pd thickness with a period of 6 atomic layers at low temperature. We ascribe the modulation of the magnetic anisotropy in Co to thickness dependent crossing of states of the Fermi energy due to QWS in Pd. We observe larger amplitude oscillations of the MA in Pd/Co as compared to the Cu/Co system. These results are additional ingredients for the understanding of electron confinement effects in spin-polarized systems and for tuning magnetic anisotropy in magnetic multilayers at the nanoscale. Further experiments on systems with even larger spin-orbit coupling of the non-ferromagnetic component, such as Pt, would be interesting, as possibly even larger anisotropy oscillations could be induced. 
 
\section{Acknowledgments}
The authors deeply acknowledge Marek Cinal (PAN, Warsaw, Poland) and Eric Montoya (UC, Riverside) for fruitful discussion. We thank W. Greie, F. Weiss and H. Menge for their technical support. 

%

\end{document}